\documentstyle[11pt]{article}
\title{Constraints on mirror fermion mixing angles\\
from anomalous magnetic moment data }

\author{
 F. Csikor and Z. Fodor\\
 Institute for Theoretical Physics, E\"otv\"os University \\
 Budapest, Hungary \\ }

\date{December 1991}
\begin{document}
\maketitle

\begin{abstract} \normalsize

The new contributions to the electron (muon) anomalous magnetic moment
arising  in mirror fermion theories have been calculated. Imposing the
experimental constraint lowers the current upper bound on the
ordinary - mirror lepton mixing angles
by a factor of 50 making predictions for mirror lepton production at
HERA undetectably small. A way out is to allow for different mixing
angles of the L and R field components. Choosing very small right
mixing angles compatibility with the anomalous magnetic moment
measurement may be easily maintained, while choosing left mixing angles
close to the upper limits   yields still reasonable HERA
cross-sections.
\end{abstract}
\newpage   \vspace{2cm}
\vspace{2cm}
\noindent
1. Mirror fermions are hypothetical particles (for a review see
\cite{MR1}) which should have opposite
chirality propeties as compared to the known ordinary, light fermions.
Thus right handed mirror fermions form weak SU(2) doublets, while the
left handed fields are singlets.
Since mirror fermions have not been observed directly one has to search
for their effects indirectly through the modification of ordinary fermion -
vector boson couplings. Such an investigation has been performed in \cite{LL},
 \cite{AC}
where the hermitian mirror fermion model of \cite{IM} has been explicitely
considered.  A recent update of essentially the same analysis
has been published in
\cite{EN}. In both analyses the authors allow for different mixing angles of
the L and R components of ordinary and mirror fermion fields. The result of
\cite{EN} is that the upper bounds on the sine squares of  leptonic mixing
angles  are around 0.10 - 0.15, while the somewhat higher bounds
for quarks are not less than 0.02.
Since \cite{EN} did not equate the L and R mixing angles the bounds valid in
the hermitian mirror fermion model may be even slightly higher.\\

Another source of information on the mixing angles  comes from higher order
electroweak corrections. A particularly sensitive observable is the anomalous
magnetic moment of the electron (muon). This quantity  has already been studied
from the point of view of mirror fermion mixing angles \cite{MR}.  However, the
range of allowed values of mirror fermion masses has changed dramatically,
therefore a reanalysis is timely. The new bounds on the masses come from LEP
experiments, restricting the number of light  neutrinos to three and imposing a
lower bound on the  mirror masses of about half the Z boson mass. If
mixing angles are not unexpectedly small even masses upto the Z mass
are excluded. Therefore we shall consider the case of heavy mirror
fermions with masses of the order of 100GeV. In order not to get into
trouble with one loop electroweak corrections \cite{KL} we shall also
assume almost degenerate doublets. In this case LEP experiments exclude
more than three heavy mirror fermion generations at the 90\% C.L. \cite{BBR}.
The tree unitarity constraints on
the mirror masses are rather low, of the order of a few hundred GeV (for
more details see \cite{FCs}.)
A new feature
of our analysis is that we have found a relation between the masses and mixing
angles, which has not been considered in \cite{MR}. Therefore we get a
completely new situation and restrictive bounds on the mixing angles, which
seriously modify mirror lepton production expectations \cite{IM2,FCs}
at future ep colliders like HERA. \\

\noindent 2. The mirror fermion model we consider has the standard
model gauge group
and symmetry breaking, the only modification being that mirror partners
of all ordinary standard model fermions are introduced. Ordinary and
mirror fermions may mix. In principle the mixing angles of the L and
R fields may be different, equating them we obtain the hermitian mirror
fermion model of \cite{IM}. The leptonic weak interaction part of the
model is written down in detail in \cite{MR}, the notations of which
will be used in the following.   Some consequences of such a model have been
worked out (especially for electron proton colliders) in \cite{FCs}.\\

The calculation of the contribution arising from mirror fermions to the
anomalous magnetic moment of the electron (muon) is straightforward. The graphs
to be taken into account are shown in Fig.1. Wherever a vector boson exchange
occurs a Goldstone boson should be also exchanged, when the calculation is
performed in a renormalizable gauge.  Our result reads:
\begin{displaymath}
\Delta\mu_e =\frac{G_F m_e \mu_B}{32\pi^2\sqrt{2}}\times
\end{displaymath}
\begin{eqnarray}
\left(M_Z\,h_1 (\frac{m_E}{M_Z})\sin^2 (2\Theta)\,-\,2M_W\,h_2(\frac{m_N}{M_W})
\sin (2\Theta)\cdot\sin (2\Phi)\right)+\Delta\mu_e^{Higgs}, \nonumber \\
\end{eqnarray}

where
\begin{displaymath}
h_1 (x)=-\frac{6x^3}{(x^2-1)^3}\cdot\ln{x^2} +\frac{x^5+x^3+4x}{(x^2-1)^2}
\end{displaymath}

\begin{displaymath}
h_2 (x)=\frac{6x^5}{(x^2-1)^3}\cdot\ln{x^2} +\frac{x^5-11x^3+4x}{(x^2-1)^2}
\end{displaymath}
and
\begin{eqnarray}
\Delta\mu_e^{Higgs} =\frac{G_F m_e \mu_B}{64\pi^2\sqrt{2}}\cdot
M_H\,h_3 (\frac{m_E}{M_H})\sin^4 (4\Theta) ,
\end{eqnarray}

\begin{displaymath}
h_3 (x)=\left(-\frac{1}{(x^2-1)}\cdot\ln{x^2}
+\frac{3-x^2}{2}\right)\cdot \frac{x^3}{(1-x^2)^2} .
\end{displaymath}
$\Theta$ and $\Phi$ are the electron - mirror electron, neutrino - mirror
neutrino
mixing angles, respectively,
$\mu_B$ the Bohr magneton. $m_e$, $m_E$, $m_N$ denote the masses of the
electron, mirror electron and mirror neutrino. Other notation is standard. \\

Note the small change as compared to the formula of \cite{MR}.
In particular the
relative sign ot the first two terms has been changed
and a misprint in the function
$h_2$ has been corrected.
The above formula applies to mirror fermions. However, in a general
theory the formula
still applies. The mixing angle factors correspond to $v^2-a^2$ of the neutral
(charged) current coupling  vertex of fermion-exotic fermion-vector boson.
Therefore a modification of the formula is straightforward. \\

As well known the standard model prediction for the anomalous magnetic
moment of the electron and muon is in good agreement with
experiment. Thus the new contribution should be small, essentially
 of the order of the experimental error.
The experimental error on electron (muon) anomalous magnetic moment is
$10^{-11} \mu_B$ ($8.1\cdot 10^{-9} \frac{e\hbar}{2m_{\mu}}$) \cite{RT}.
With the usual values of the mixing angles (i.e. $\sin^2 \Theta ,\: \sin^2
\Phi \, \approx 0.02$) one gets
much larger values than the experimental error. The Higgs contribution is small
so it can be safely neglected.  The only way to achieve a reasonable value is
to allow for the cancellation of the two terms containing $h_1$ and $h_2$. That
would lead to a relation between the two mixing angles. On the other hand, a
detailed investigation of the fermion mass generation via the usual
electroweak symmetry breaking shows that the mixing angles can not be choosen
arbitrarily.  We get the relation
\begin{eqnarray}
-m_{\nu} \sin \Phi \cos \Phi +m_N \cos \Phi \sin \Phi \,=\,
-m_{e} \sin \Theta \cos \Theta +m_E \cos \Theta \sin \Theta, \nonumber \\
\end{eqnarray}
which makes the necessary cancellation of the two terms possible
only if the N,
E mass splitting is unreasonably large. (E.g. for exact cancellation $m_N /m_E
\approx 13$.)
  Fixing the value
of $\Delta \mu _e$ to $10^{-11} \mu _B$, and using Eqs. (1,3)  one gets the
curves of Fig.2, specifying the mixing angles and masses which yield acceptably
small contribution to the anomalous magnetic moment. (The mass of the mirror
electron neutrino is determined by Eq.(3).) We see that the mixing
angles have to be very small, with the square of the order of 0.0004.
This corresponds to $m_E \approx m_N$ as required by consistency with the
LEP results \cite{KL,BBR}.
(The muon anomalous magnetic moment error constrains the mixing angles
of the second generation lepton mixing angles in a similar way,
with approximately the same upper limit.)
This is a factor 1/50 lower that the values assumed in \cite{FCs}.
Since the HERA production cross-sections are proportional to the mixing
angle squares, the predictions of \cite{FCs} decrease roughly by a
factor of 1/50,  making mirror lepton productions at HERA assuming the
usual $100pb^{-1}$ yearly integrated luminosity essentially hopeless. \\

Let us now consider the
case of a mixing scheme with different L and R mixing angles. In this
case the following
replacements should be made in Eq.(1).
\begin{displaymath}
\sin (2\Theta )\sin (2\Phi )\longrightarrow 4\cos \Theta_L \sin \Theta_R \cos
\Phi_R \sin \Phi_L
\end{displaymath}
\begin{eqnarray}
\sin^2 (2\Theta )\longrightarrow \sin (2\Theta_R )\sin (2\Theta_L) .
\nonumber \\
\end{eqnarray}
{}From the structure of the symmetry breaking
mechanism one gets the relation:
\begin{eqnarray}
-m_{\nu} \sin \Phi_R \cos \Phi_L +m_N \cos \Phi_R \sin \Phi_L \,=\,
-m_{e} \sin \Theta_R \cos \Theta_L +m_E \cos \Theta_R \sin \Theta_L
\nonumber \\
\end{eqnarray}
Neglecting $m_{\nu}$ and $m_e$ one gets a relation sensitive only to
the L mixing angles. Neglecting the Higgs contribution and inserting into
 Eq.(1) we get :

\begin{displaymath}
\Delta\mu_e =\frac{G_F m_e \mu_B}{32\pi^2\sqrt{2}}\times
\end{displaymath}
\begin{eqnarray}
\sin (2\Theta _R)\sin (2\Theta _L)
\left(M_Z\,h_1 (\frac{m_E}{M_Z})\,-\,2M_W\,h_2(\frac{m_N}{M_W})
\cdot \frac{m_E}{m_N}\right). \nonumber \\
\end{eqnarray}
This shows that either the R or the L angle has to be small in order to
suppress the contribution. Generalizing the formulae of \cite{FCs} we may
write in the small mixing angle approximation the HERA mirror lepton
production cross-sections:

\begin{eqnarray} \label{eq06}
\lefteqn{\frac{d\sigma_{e^-u \rightarrow N_e d}}{dQ^2} \simeq
\frac{g^4}{64\pi(M_W^2 + Q^2)^2} \times } \nonumber \\
 & & \left[
\Phi _L^2 \left( 1 - \frac{M^2}{xs} \right) +
\Theta _R^2 (1-y) \left( 1-y-\frac{M^2}{xs} \right)
\right] \nonumber \\
\end{eqnarray}
\begin{eqnarray} \label{eq07}
\lefteqn{\frac{d\sigma_{e^-q_A \rightarrow E^-q_A}}{dQ^2} \simeq
\frac{(g^2 + g^{\prime 2})^2}{1024\pi(M_Z^2 + Q^2)^2} \cdot } \nonumber \\
 & & \left[
 \Theta _L^2 \left( (1-4\cdot \mid Q_A \mid \sin ^2 \Theta _W)^2
(1-\frac{M^2}{xs})+ (4Q_A ^2\sin ^2 \Theta _W )^2 (1-y)(
1 - y - \frac{M^2}{xs})\right) \right.+\nonumber \\
 & & \left.
\Theta _R^2 \left( (1-4\cdot \mid Q_A \mid \sin ^2 \Theta _W)^2 (1-y)
(1-y-\frac{M^2}{xs})+ (4Q_A ^2\sin ^2 \Theta _W )^2 (
1 - \frac{M^2}{xs})\right)
\right] \nonumber \\
\end{eqnarray}
We see\footnote{This possibility has been pointed out to us by I. Montvay.}
that choosing $\Theta _R$ to be small both the mirror neutrino
and mirror electron cross-sections have an unconstrained L mixing angle
factor. Assuming the \cite{EN} upper limits we get the numbers of Table
1, which are only about a factor 2 lower than the predictions presented
in \cite{FCs}.  \\

\noindent 3. Let us now discuss how Eqs. (3,5) follow from the
electroweak symmetry
breaking. It is sufficient to consider the first family ordinary and
mirror leptons. Assuming a standard scalar sector the most general
$SU(2)\otimes U(1)$ Yukawa couplings
and mass mixing terms of the Lagrangian may be written as follows.
Introduce the notation
\begin{displaymath}
\psi _L ^o =\left( \begin{array}{c}\nu _L ^o \\e_L ^o \end{array}\right),\:
\chi _R ^o =\left( \begin{array}{c}N _R ^o  \\E_R ^o \end{array}\right),\:
\end{displaymath}
\begin{displaymath} e_R ^o ,\: \nu _R ^o ,\: N_L ^o ,\: E_L ^o ,
\end{displaymath} 
\begin{eqnarray}
\phi  =\left( \begin{array}{c}\phi ^o \\ \phi ^-\end{array}\right),\:
\tilde \phi  =\left( \begin{array}{c}\phi ^+ \\ -\phi ^{o*}\end{array}\right).
\end{eqnarray}
Then
\begin{displaymath}
{\cal L}_{Yukawa}=g_1 \overline{\psi _L ^o} \tilde{\phi} e_R ^o-
g_1 ^{\prime } \overline{\psi _L ^o} \phi \nu_R ^o
+g_2 \overline{\chi _R ^o} \tilde{\phi} E_L ^o-
g_2 ^{\prime } \overline{\chi _R ^o} \phi N_L ^o
\end{displaymath}
\begin{equation}
+g_3 \overline{\chi _R ^o}\psi _L ^o +g_4 \overline{e_R ^o} E_L ^o+
g_5 \overline{\nu _R ^o} N_L ^o +h.c.
\end{equation}
Besides the Yukawa-couplings we have as well included the $SU(2)
\otimes U(1)$ invariant contact terms responsible for the
ordinary - mirror fermion mixing.
Inserting the vacuum expectation value of the scalar field, we get as usual
the mass matrices:
\begin{displaymath}
(\overline{e_R ^o},\overline{E_R ^o})\left( \begin{array}{cc}
-g_1 v/\sqrt{2} &g_4 \\ g_3 &-g_2 v/\sqrt{2}\end{array} \right) \left(
\begin{array}{c} e_L ^o \\ E_L ^o \end{array} \right) \:+\:h.c.
\end{displaymath}
\begin{equation}
(\overline{\nu_R ^o},\overline{N_R ^o})\left( \begin{array}{cc}
-g_1 ^{,} v/\sqrt{2} &g_5 \\ g_3 &-g_2 ^{,} v/\sqrt{2}\end{array}
\right) \left(
\begin{array}{c} \nu _L ^o \\ N_L ^o \end{array} \right) \:+\:h.c.
\end{equation}
Assuming CP conservation $g_4 =g_4 ^*$, $g_5 =g_5 ^*$ and $g_3 =g_3 ^*$
 we can diagonalize with real unitary matrices (i=L,R)
\begin{displaymath}
\left(  e_i ^o ,E_i ^o \right) =\left( \begin{array}{cc} \cos \Theta _i & \sin
 \Theta _i \\ -\sin \Theta _i & \cos \Theta _i \end{array} \right)
\left( \begin{array}{c} e_i \\ E_i \end{array} \right)
\end{displaymath}
\begin{equation}
\left(  \nu _i ^o ,N_i ^o \right) =\left( \begin{array}{cc} \cos \Phi _i & \sin
 \Phi _i \\ -\sin \Phi _i & \cos \Phi _i \end{array} \right)
\left( \begin{array}{c} \nu _i \\ N_i \end{array} \right)
\end{equation}
The relations
\begin{displaymath}
g_3 =-m_e \sin \Theta _R \cos \Theta _L + m_E \sin \Theta _L \cos \Theta _R
\end{displaymath}
\begin{equation}
g_3 =-m_{\nu} \sin \Phi _R \cos \Phi _L + m_N \sin \Phi _L \cos \Phi _R ,
\end{equation}
arise from the electron and neutrino mass matrices, respectively.
Equating the two expressions of $g_3$, we get Eq.(5), mentioned before. The
origin of the relation is that the $g_3 \overline {\chi _R ^o} \psi _L ^o +
h.c.$  term makes a connection between the parameters of the electron and
neutrino mass matrices. \\

Excluding $\nu _R^o$ from the initial fields we get instead of Eq. (13)
\begin{displaymath}
g_3=m_N \sin \Phi _L ,
\end{displaymath}
which  coincides with Eq. (13) provided  $m_{\nu} =0,$
$\Phi _R =0$ are choosen.
Since $m_E ,\: m_N \gg m_e ,\: m_{\nu} $ and the mixing angles are small,
this possibility does not much differ from the previous one.
It is clear that similar relations do arise for all particle pairs, whose
L components form a doublet and have mirror partners. \\

\noindent
4. We have studied the constraints on mirror fermion mixing angles
arising from
a comparison with anomalous magnetic moment data. We have found a
new relation Eq.(3) (or Eq.(5)) connecting mirror mixing angles and fermion
masses. Taking  this relation into account makes reconciliation of anomalous
magnetic moment data with not too small mixing angles difficult in the case
when left and right mixing angles are equal. An appealing possibility leading
to reasonably large mirror lepton production cross-sections in electron proton
collisions is to assume a very small right electron mixing angle, with left
mixing angles close to the experimental limits of \cite{LL,EN}. \\

\noindent {\bf Acknowledgements.} We thank W. Bernreuter, W. Grimus
and I. Montvay for helpful discussions. \\

\vspace{1cm}

\newpage
\noindent {\bf Table 1.} Total production cross-section
$\sigma $ of the mirror electron-neutrino and
mirror electron in electron-proton collisions at
$\sqrt s$=314GeV as a function
of the mirror lepton mass. Left mixing angle squares are assumed
to be 0.02, while the right
mixing angles are zero. The cross-sections are in $10^{-2} pb$, masses are
in GeV. \\

\vspace{2cm}
\begin{center}
\begin{tabular}{ccc}
 M & $\sigma (ep\rightarrow NX)$ & $\sigma (ep\rightarrow EX)$ \\
 100 & 19.5 & 4.7 \\
 120 & 11.8 & 2.7 \\
 140 & 6.8 & 1.5 \\
 160 & 3.5 & 0.7 \\
 180 & 1.7 & 0.3 \\
 200 & 0.7 & 0.13
\end{tabular}
\end{center}

\newpage
\vspace{1cm}
\noindent {\bf Figure Captions} \\

\vspace{1cm}
\noindent {\bf Fig. 1.} Diagrams for the calculation of the mirror fermion
contribution to the
magnetic moment of the electron.  \\

\noindent {\bf Fig. 2.} Curves showing the possible values of the
mixing angles for different mirror electron  masses. The value of the mirror
fermion contribution to the anomalous magnetic moment of the electron
is set equal to the experimental error.
\end{document}